\documentclass[11pt,aps,prb,preprint]{revtex4}
\usepackage{amsfonts,amsmath,amssymb,graphics,epsfig}
\usepackage{enumerate}
\usepackage{theorem}

\begin{document}
\title{ {\large {\bf  Imperfect fluids, Lorentz violations and Finsler Cosmology }}   }

\author{A.P.Kouretsis }
\affiliation{Section of Astrophysics, Astronomy and Mechanics, Department of Physics  Aristotle University of Thessaloniki, Thessaloniki 54124, Greece}
\email{akouretsis@astro.auth.gr}

\author{M.Stathakopoulos}
\affiliation{1 Anastasiou Genadiou Street, 11474, Athens, Greece}
\email{michaelshw@yahoo.co.uk}

\author{P.C.Stavrinos}
\affiliation{Department of Mathematics University of Athens 15784 Greece}
\email{pstavrin@math.uoa.gr}

\begin{abstract}
We construct a cosmological toy model based on a Finslerian structure of space-time. In particular, we are interested in a specific Finslerian Lorentz violating theory based on a curved version of Cohen and Glashow's Very Special Relativity. The osculation of a Finslerian manifold to a Riemannian  leads to the limit of Relativistic Cosmology, for a specified observer. A modified flat FRW cosmology is produced. The analogue of a zero energy particle unfolds some special properties of the dynamics. The kinematical equations
of motion are affected by local anisotropies. Seeds of Lorentz Violations may trigger density inhomogeneities to the cosmological fluid.
\noindent {\small  \newline\newline {\bf Key
words}:Finsler Geometry, Cosmology,  Very Special Relativity, Lorentz violations}\newpage
\end{abstract}

\maketitle

\section{Introduction}

The idea of Lorentz Violations (LV) in modern physics can be traced back to the studies of Dirac in late 50's  \cite{hist}. Contemporary research problems of high energy physics lead to the formulation of various quantum gravitational (QG) theories  which inherit local anisotropies in most circumstances. A sensible approach to this direction is the study of GR's extensions, where new space-time symmetries are introduced. Noteworthy formalizations in this direction, are extra dimensional physics  and the non-commutativity of space-time geometry (see e.g. \cite{Maartens:2003tw,noncom}). In this framework, new physical phenomena emerge in long range distances that may resolve questions of modern theoretical physics. A common feature in QG is the prediction of a modified mass-shell condition for elementary particles.  Moreover, these departures from Lorentz invariance predict a vacuum refractive index and corrections at the threshold energy. The most debated effects  are the  time delay  of light-rays which depends on the energy of photons (see e.g. \cite{AmelinoCamelia:1997gz}), and threshold anomalies reported from astrophysical observations \cite{AmelinoCamelia:2008qg}.

This phenomenology was recently associated with a velocity dependent geometry called Finsler.  In particular, F.Girelli et.al \cite{Girelli} argued  that several structures, like the ``rainbow metric'' \cite{Magueijo:2002xx} and other alternative scenarios of Deformed Special Relativity-like \cite{AmelinoCamelia:2000mn} models, can be approached by a Finslerian perspective. The same formalism seems to be compatible with the propagation of rays in Horava-Lifshitz gravity \cite{Sindoni:2009vj}. Another case where Finsler-like structures appear is the D-particle recoil example, where the effective 4-dimensional metric depends on phase space coordinates \cite{mavr,Li:2009tt,Mavromatos:2007sp}. Similar scenarios from a different point of view on string-like theories have been discussed   in \cite{Vacaru:general}. We also mention the correlation of birefringence optics to Finslerian space-times \cite{Visskal}. The aforementioned phenomenology suggests that Finsler geometry could play a fundamental role in modern QG theories.

Another case on Lorentz violations is the minimalistic approach of Very Special Relativity (VSR). The construction of VSR is based on a proper subgroup of the Poincare group. An induced ``{\it ether}'' moving with the speed of light simulates a null spurionic vector field. In the context of VSR the introduction of a neutrino mass requires no additional states and needs no violation of leptonic number \cite{Coh}. However, departures from Special Relativity and CPT invariance are difficult to detect due to the null nature of ``{\it ether}'' \cite{Dunn:2006xk}. The above construction is incorporated to the Finslerian framework again, after considering a curved version of VSR namely General Very Special Relativity (GVSR) \cite{Gibb_fins}.

GVSR is manipulated to build  a cosmological toy model. We use a  similar approach to \cite{kourPRD}, with the essential difference that the null character of the spurion is preserved in alliance to the original GVSR theory.  A  better insight about the effects of Finsler Geometry to gravitational physics is achieved by the osculation of a Finslerian manifold to a Riemannian one. We consider that the physical geometry is represented by a Finslerian space-time, while gravitational geometry is described by a Riemann structure, following \cite{bekenstein}. Therefore, the FRW metric is invoked to the Finslerian background to study deviations from the standard cosmological picture. In this framework,  an observer falls on  a peculiar non-geodesic congruence with respect to the FRW comoving motion.

This paper is organized as follows: After a short introduction to Finsler Geometry we outline the model of GVSR and some of its phenomenological consequences. In Sec.3 we describe the useful tool of the osculating Riemannian space and its implications to gravitational physics. In Sec.4 we apply the osculating process to the model of GVSR using the FRW metric for a flat universe. Sec.5,6 are devoted to the construction of the equations of motion and continuity, relied on the induced space-time symmetries. In Sec.7 we present the solutions for the scale factor, flux and anisotropic pressure, while in Sec.8 we discuss the implied long range modifications depicted by a modified FRW potential. Besides, we attempt to relate the dynamical behavior to some large scale observables. It seems that exotic matter or lower values for the energy density of the cosmic fluid is required to generate late time acceleration. Sec.9,10 further analyze the kinematical properties of the model, to achieve some level of insight about the evolution of the medium. Finally, we highlight for future development, that  departures from the safe harbor of Riemann geometry may trigger density perturbations, leaving artifacts of Lorentz violations.

\section{Finsler Geometry and Very Special Relativity}
 Finsler geometry is a generalization of Riemann geometry, where all the geometrical structures depend on the element $(x^i,\dot{x}^i)$ rather than the position coordinate solely. The line element is defined by a norm $F(x,\dot{x})$ over the tangent bundle $TM\backslash \{ 0 \}$, where $M$ is the base manifold. The $F(x,\dot{x})$ is a homogenous function of first degree with respect to $\dot{x}$, such that the integral of the arc-length $\int F(x,\dot{x})d\tau$ is independent of the parameter $\tau$. Finsler geometry is strictly discriminated from Riemann geometry after dropping the quadratic restriction over the metric function $F(x,\dot{x})$ \cite{RundFB} i.e.
 \begin{equation}
 F^2(x,\dot{x})=f_{\mu \nu}(x,\dot{x})\dot{x}^\mu \dot{x}^\nu \label {Fsq} .
 \end{equation}
 Using Euler's theorem we can calculate the Finsler metric
 \begin{equation}
f_{\mu\nu}(x,\dot{x})=\frac{1}{2}\frac{\partial^2 F^2}{\partial \dot{x}^\mu \partial \dot{x}^\nu}(x,\dot{x})\label{fmn}
\end{equation}
which is homogeneous of zero degree with respect to $\dot{x}$.
  The definition (\ref{fmn}) is reduced to a Riemannian metric when the metric tensor depends solely on the position $x^i$,  indicating that (\ref{Fsq}) is a quadratic form in $\dot{x}^j$. Therefore, Finsler geometry can be considered as a natural generalization of Riemann geometry.

 The unit sphere $I_x=\{F(x,\dot{x})=1, \;\dot{x}\epsilon T_xM\}\subset T_xM$  is called the {\it indicatrix}  and defines a 3-dimensional locus in every tangent space $T_xM$. In case of a Riemannian space the indicatrix is an ellipsoid as a result of the quadratic restriction. However, (\ref{Fsq}) implies that the tangent spaces of a Finsler space are not equipped with  ellipsoidal unit balls as in Riemann geometry, generating local anisotropies of space-time. Therefore a geometrical property should arise to describe this ``distortion'' of the indicatrix, called color (see e.g. \cite{Shen}). In particular, a Riemannian space is considered entirely ``white'', while in most cases a Finsler space possesses different color patterns over the manifold.

 The lack of quadratic restriction  appears in some Phenomenological Quantum Gravitational theories as a consequence of Lorentz symmetry breaking \cite{Girelli}-\cite{ Visskal}. Thus, in such a scenario,  quantities which measure the color and its variations, are directly related to Lorentz violations. These space-times can be characterized  as  {\it colorful curved manifolds } providing a way to study gravitational phenomena under the hypothesis of Lorentz violations \footnote{In general, Finslerian manifolds are colorful curved manifolds. The Cartan torsion $C_{\mu\nu\rho}$ and the $S$-curvature measure the color and its variations respectively.   }.

An intriguing case of Lorentz symmetry breaking where Finsler Geometry turns up, is Cohen and Glashow's Very Special Relativity (VSR) \cite{Coh}. The Lorentz violations are generated by the $ISIM(2)$ subgroup of Lorentz transformations.
Gibbons et.al investigated a deformation of VSR called {\it General Very Special Relativity} (GVSR) \cite{Gibb_fins}. Among the deformations of $ISIM(2)$ there is an 1-parameter family called the $DISIM_b(2)$. This deformation group leaves invariant the Finslerian line element,
\begin{equation}
ds=(\eta_{ij}dx^idx^j)^{(1-b)/2}(n_kdx^k)^b\label{DS}
\end{equation}
proposed by Bogoslovsky (see \cite{Bogoslovsky} and references there in) for the study of local anisotropies. The entity  $n^k$ is a  null spurionic vector field  that determines the direction of the ``etheral'' motion's 4-velocity and can be selected as $n^i=(1,0,0,1)$ \cite{Coh, Gibb_fins, Bogoslovsky:2004rp}. The
signature of the Minkowski metric is set to $\eta_{ij}=\mbox{diag}(+1,-1,-1,-1) $ through-out this paper.  The line element (\ref{DS})  determines  a particle's mass-tensor
\begin{equation}
m_{ij}=(1-b)m\left( \delta_{ij}+bn_in_j \right), \label{mij}
\end{equation}
which indicates the Machian nature of the theory \cite{Bogoslovsky}.

The symmetry of the line element (\ref{DS}) indicates the Lagrangian of a free-moving particle
\begin{equation}
\mathcal{L}=m(\eta_{ij}\dot{x}^i\dot{x}^j)^{(1-b)/2}(n_k\dot{x}^k)^b. \label{Lg}
\end{equation}
The above Lagrangian implies the particle's action integral $I=\int \mathcal{L}(x,\dot{x})d\tau$ and the canonical momentum $ p_k=\frac{ \partial \mathcal{L} }{ \partial \dot{x}^k }$ \footnote{Note that $p_\mu=m\left[ \left( 1-b \right) L^{\frac{-1-b}{2}}N^{b}y_\mu+bL^{ \frac{1-b}{2} }N^{b-1}n_\mu  \right]$ reveals the effect of the preferred direction on the particle's dynamics. The scalars $L,N$ are defined in (\ref{fmn_explc}). }. Using the first order degree homogeneity of the Lagrangian with respect to $\dot{x}$, we can construct the generalized mass-shell condition
\begin{equation}
f_{\mu\nu}(x,\dot{x})p^\mu p^\nu=m^2
\label{fmscd}
\end{equation}
since the tangent and cotangent bundles define equivalent geometrical frameworks \cite{Girelli}. Condition (\ref{fmscd}) can be reformed to the following convenient expression
\begin{equation}
\eta^{ij}p_ip_j=m^2(1-b^2)\left( \frac{n^ip_i}{m(1-b)} \label{ndr}      \right)^{2b/(b+1)}.
\end{equation}
This relation  reflects  the colored nature of the space-time manifold. The parameter
$b$ is  restricted by various experiments to the order of $|b|< 10^{-10}$  ({\it ether drift experiment}) and
$|b|< 10^{-26}$ {\it (anisotropy of inertia)} \cite{Gibb_fins,Bog1}.

The metric  (\ref{fmn}) of Finsler geometry does not  describe completely the geometric properties of the underlying  manifold as in the Riemannian case.  Further information must be supplied concerning the (non) linear connection. We refer two main branches of formalism among others, characterized by Cartan's and Chern's connections. The Chern's approach introduced a connection which gave a complete system of local invariants ensuring that two Finsler structures differ by a change of coordinates (see for example \cite{Shen}). Nevertheless, it is an almost metrical ansatz. On the other hand, Cartan's connection is purely metrical. Physical implications of the aforementioned  perspectives, suggest a non-conservation of energy and momentum apart from a subclass of Finsler spaces called Berwald spaces (see e.g. \cite{Vacaru:general,ChGen} and references therein). However, it is rather obscure if departures from Relativistic invariance anymore guarantee the energy-momentum conservation and/or metricity of space-time. In the following sections we use a method of great simplicity, initially developed for the purpose of comparing various covariant derivatives of Finsler Geometry, {\it the process of osculation}. In this approach, a purely Riemannian metric is defined in a local subregion (see next section) and GR's machinery is valid for the induced Levi-Civita connection.

Before we proceed to the limit of osculation we investigate the mass-shell condition (\ref{ndr}) which may provide observational motivations as we expect  modified dispersion relations (MDR)  and an influence on  threshold energies. These phenomenological concepts  drew some attention due to the possibility of constraining various QG models with current astrophysical data ( time of flight differences between photons of different energies,  TeV-$\gamma$ and Ultra High Energy Cosmic Rays (UHECR) threshold anomalies) \cite{AmelinoCamelia:2008qg}. In order to demonstrate some  physical implications of the modified mass-shell condition (\ref{ndr})  we will roughly compute the threshold energy of $p+\gamma\rightarrow p+\pi$  for the Finslerian background (\ref{DS}). The threshold anomalies have also been studied for a Finsler-Randers space \cite{ChinThr}. For the sake of simplicity, we will assume energy and momentum conservation
\begin{equation}
E_1+\epsilon=E_2+E_3,\:\: p_1-q=p_2+p_3
\end{equation}
where $E_1$ and $p_1$ refer to a high energy particle colliding with a photon $(\epsilon,q)$ and $E_2,p_2,E_3,p_3$ represent the energy and momentum of the produced particles. The non-Lorentz invariant relation (\ref{ndr}),  for a further assumption of massless photons and small departures from Lorentz invariance, reads for a photon and the $i$-particle,
\begin{equation}
q=\epsilon\:,\:E_i^2-p_i^2=m_i^2\left[1-2\ln\left(\frac{E_i-p_i\beta_i}{m_i}\right)b+O(b^2)\right]
\label{MRLth}
\end{equation}
where $\beta_i=\cos\theta_i$ and $\theta_i$ denotes the angle between the particle's spatial momentum and the spatial part of the null spurion. The dependence of (\ref{MRLth}) on the parameter $\beta_i$, reflects  the local anisotropic structure of the Finslerian space-time. The same particle with different orientation  possesses a different energy component. Working only on the lab frame,  we can side-step the definition of different local frames in our LV-theory \cite{AmelinoCamelia:2008qg}. The limit of high energy particles leads to the following formula for the threshold energy
\begin{equation}
E_1\simeq\frac{(m_2+m_3)^2-m_1^2}{4\epsilon}-\frac{(m_2+m_3)^2}{2\epsilon}\ln\left[\frac{E_1(1-\beta)+\epsilon(1+\beta)}{m_2+m_3} \right]b\label{thres}
\end{equation}
where $\beta$ stands for the angle between the total 3-momentum of the produced particles and the spurion's spatial preferred direction. We remark that, the LV effect to the threshold energy fades out parallel to the spatial direction ($\beta=1$).  The above particle's production is expected between  soft photons of the CMB and UHECRs. In general, we don't expect to observe  UHECRs above the well-known GZK cutoff  $\sim 5\cdot 10^{19}eV$  \cite{obspapAC}. Particles with energy values above this cut-off should be absorbed by the CMB background. However,  cosmic rays with energies up to $\sim 3\cdot 10^{20}eV$ have been reported. A possible explanation is that the threshold of the photonpion process is at larger energies. In that spirit we can constrain various parameters coming from QG scenarios. For our case the value of $b$ is constraint up to $b\sim-0.1/(27.7+\ln(1-\beta))$. This is not a tight constraint for our parameter but it clearly depicts that locally a Finsler space is directional dependent. Nevertheless, the reader should keep in mind that energy-momentum conservation is not necessarily valid in a Finslerian background. Also, non trivial physics is involved due to the angle-dependence of the mass-shell condition. Last but not least, a proca-like Lagragian is invariant under $DISIM_b(2)$, therefore massive photons should be embodied to the calculations.

\section{The process of osculation}
Let $U$ be a region of the space-time Finsler manifold $F_4$ parameterized by the local coordinates $x^\mu$. In this neighborhood the velocity vector field $u^\mu$ can be picked up to be $x^\mu$ dependent, $\dot{x}^\mu=u^\mu(x)$.
Thus this local region of $F_n$ possesses a Riemannian metric
\begin{equation}
g_{\mu\nu}(x)=f_{\mu\nu}(x,u(x))\label{oscmetr}
\end{equation}
and leads to the construction of the {\it Osculating Riemannian Space}. In other words, the process of osculation relates the velocity field uniquely to the space-time points, implying the limit of relativistic cosmology \cite{raych,raych2}.
The osculation of a Finslerian manifold determines a purely Riemannian space-time for a specified selection of $u^\mu(x)$. In case of a general fluid the comoving observers live on a different Riemann space to the one of a tilted observer. In such a scenario the link between different families of observers belongs to the context of Finsler geometry \cite{Ish,Asanov}.

Consider an early time of the universe where Lorentz violations and chaotic motion govern the cosmic fluid.  Finsler geometry is a candidate physical geometry of this state. As the universe evolves, Lorentz violations and chaotic motion are expected to fade out, enabling the introduction of the observer's 4-velocity with respect to space-time coordinates. Hence, the definition of an osculating Riemannian space is possible, using the observer's 4-velocity field. A post big-bang Riemannian cosmology rises up where seeds of the Finslerian era can survive.
The induced Einstein field equations for the $u^\mu$ observer are
 \begin{equation}
 G_{\mu \nu}(x,u(x))=8\pi GT_{\mu \nu}(x,u(x))\label{fieldequat}
 \end{equation}
where $G_{\mu \nu}$ is the standard Einstein tensor coming from the Riemann metric (\ref{oscmetr}). The energy-momentum tensor represents a general imperfect fluid  which can be expressed into its irreducible parts (for a recent review see \cite{tsagas_rep})
\begin{equation}
T_{\mu\nu}=\mu u_\mu u_\nu-Ph_{\mu\nu}+2q_{(\mu}u_{\nu)}+\pi_{\mu\nu}
\label{Tmn}
\end{equation}
where $h_{\mu\nu}=g_{\mu\nu}-u_\mu u_\nu $ represents the projective tensor, $\mu=T_{\alpha\beta}u^\alpha u^\beta$ is the energy density, $P=-T_{\alpha\beta}h^{\alpha\beta}/3$ is the isotropic pressure
of the fluid coming from the equilibrium pressure and bulk viscosity, $q_\mu=h_\mu^\rho T_{\rho\sigma}u^\sigma$ is the total energy flux vector and $\pi_{\alpha\beta}=h_{<\alpha}^{\:\: \gamma}h_{\beta>}^{\: \:\delta}T_{\gamma\delta}$ is the symmetric, trace-free anisotropic stress tensor \footnote{Angled brackets denote the projective symmetric and trace-free (PSTF) part of a tensor $h^\gamma_{\langle\alpha}h^{\delta}_{\beta \rangle}T_{\gamma\delta}=
h^\gamma _{( \alpha } h^{\delta}_{\beta )} T_{\gamma\delta}-\frac{1}{3} h^{\gamma\delta}T_{\gamma\delta} h_{\alpha\beta} $}.

\section{Osculating General Very Special Relativity}
We can construct our cosmological geometrical machinery by introducing the metric function
$F(x,u)$ directly from the line-element (\ref{DS}). All tensorial items of Finsler geometry obey the linear transformation law $B_{\mu}=\frac{ \partial x^i}{ \partial x^\mu }B_i$
relating different coordinate systems. The metric function $F(x,u)$ in any coordinate system is rewritten as,
\begin{equation}
\begin{array}{rl}
F(x,u)= & \left( \eta_{ij}\frac{\partial x^i}{\partial x^\mu} \frac{\partial x^j}{\partial x^\nu }u^\mu u^\nu \right)^{b-1} ( n_i
\frac{ \partial x^i }{ \partial x^\rho } u^\rho )^b \\
    = & \left( a_{\mu\nu }u^\mu u^\nu\right)^{b-1}( n_\rho u^\rho)^b
\end{array}\label{curvF}
\end{equation}
where $a_{\mu\nu}$ denotes a Riemannian metric and Greek indices represent an arbitrary coordinate system. In general the parameter $b$ can be considered $x-$dependent ($b\equiv b(x)$) preserving the 1-homogeneity of (\ref{curvF}) with respect to $u$.   The groups that leave (\ref{curvF}) invariant for different values of $b$ are not isomorphic,  leading to  geometrical phase transitions \cite{Gibb_fins,Bog1}.

The substitution of relation (\ref{curvF}) into (\ref{fmn}) implies the Finslerian metric tensor to the explicit form
\begin{equation}
\begin{array}{rl}
f_{\mu\nu}(x,u)= & (1-b)L^{-b}N^{2b}a_{\mu\nu}-2b(1-b)L^{-b-1}N^{2b}u_\mu u_\nu + \\
               + & 2b(1-b)L^{-b}N^{2b-1}n_{( \mu}u_{\nu )}+b(2b-1)L^{1-b}N^{2b-2}n_{\mu} n_{\nu}
\end{array}\label{fmn_explc}
\end{equation}
where $L=a_{\mu\nu}u^\mu u^\nu$ and $ N=n_\rho u^\rho $. We remark that (\ref{fmn_explc}) is a disformal relation between the two metric tensors $f_{\mu\nu}$ and $a_{\mu\nu}$.  Note that for a Lorentz Violating background, $f_{\mu\nu}$ plays the role of physical geometry while $a_{\mu\nu}$ represents the gravitational ``potential'' \cite{bekenstein}.

We construct a cosmological model by inserting  a flat FRW metric and the observer's 4-velocity $u^\mu=(1,0,0,0)$ into (\ref{fmn_explc}).  If Lorentz Violations  ``dilute'' to thermal energy and entropy, this set up recovers the classical FRW-limit. The process of osculation leads to a Riemannian space-time, where the observer's rest frame  lies on a tilted non-geodesic congruence.

The cosmological model is investigated during a geometrical phase where $b$ is a constant of first order disturbance upon the FRW metric. Therefore,  the osculating Riemannian line-element, by virtue of (\ref{curvF}), is reduced to
\begin{equation}
\begin{array}{rl}
ds^2=& dt^2-a^2(t)\left[ \left(1-b\right) (dx^2+dy^2)-(1-b+b/a^4(t) )dz^2 \right] \\
\: & \: \\
    \:& + 2b/a(t)dtdz
\end{array}\label{line_elem}
\end{equation}
where the null spurion has been transformed to $n^{\mu}=(1,0,0,1/a(t))$. The form of the metric (\ref{line_elem}) indicates that the background geometry includes anisotropy, since the space-time expansion is of different rate at different directions. Due to the independence of the metric components to the spatial coordinates, all the invariant quantities are only functions of time.  The off-diagonal terms of (\ref{line_elem}) imply that the $u^\mu=(1,0,0,0)$ observer will measure an energy flux component. This effective peculiar motion with respect to the FRW limit is owed to  a 4-acceleration vector $A^\mu =u^\nu\nabla_\nu u^\mu$ of Lorentz Violating origin.

Since (\ref{line_elem}) is of Riemannian nature we use the standard formulas for the connection and curvature. The Ricci tensor is directly calculated as
\begin{equation}
\begin{array}{rl}
R_{00}= &- 3\,\frac { \ddot{a }  }{a }+2a^{-5}\ddot{a}b-6a^{-6}\dot{a}^2b  \\
\: & \: \\
R_{11}=& R_{22}=   a\ddot{a}(1-b)+2\dot{a}^2(1-b)-(2a^{-4}\dot{a}^2+2\dot{a}^2+a\ddot{a})b  \\
\: & \: \\
R_{33}= & a\ddot{a}(1-b)+2\dot{a}^2(1-b)+\left(4a^{-4}\dot{a}^2-a^{-3}\ddot{a}  \right)b  \\
\: & \: \\
R_{03}= & -\frac{b}{a^3}(a\ddot{a}+2\dot{a}^2)
\end{array}\label{riccit}
\end{equation}
The Einstein tensor $G_{\mu\nu}$, combined to the field equations (\ref{fieldequat}), recasts the anisotropic irreducible parts of the energy momentum tensor (\ref{Tmn}) to the following form
\begin{equation}
q_{\mu}\propto h_{\mu}^{\:\:\alpha}R_{\alpha\beta}u^{\beta},\:\:\pi_{\mu\nu}\propto h_{<\mu}^{\:\:\alpha}h_{\nu>}^{\:\:\beta}R_{\alpha\beta}.
\label{fl_apr}
\end{equation}
The flux and the anisotropic pressure (\ref{fl_apr}), expressed in terms of purely geometrical quantities, reflect the General Relativistic interpretation of Gravity where space-time curvature determines the motion of matter.

\section{Modified Friedmann equations}
 The osculating Riemannian approach defines a cosmological {\it toy model} for a tilted observer, enriching the picture of the ''standard'' cosmology. The presence of flux, anisotropic pressure and peculiar motion reflects the assumed Finslerian background. This process relates the generated Lorentz Violations to large scale structure dynamics. Using the expressions (\ref{fl_apr}), we introduce a relativistic total fluid (\ref{Tmn}) in alliance to the anisotropic metric (\ref{line_elem})
\begin{equation}
q_\mu =(0,0,0,Q(t)),\:\: \pi_{\mu}\:^\nu =\mbox{diag} \left(0,\Pi(t),\Pi(t),-2\Pi(t) \right).
\end{equation}
where $Q(t),\Pi(t)$ are considered unknown functions. For weak deviations from the FRW cosmology $Q(t),\:\Pi(t)$ can be of first order \cite{Mart_CMB}.

The standard calculation of the Einstein field equations (\ref{fieldequat}) for a general fluid leads to the following equations of motion for the scale factor
\begin{equation}
\frac{\dot{a}^2}{a^2}-\frac{4}{3}a^{-6}\dot{a}^2b =\frac{8\pi G}{3}\mu \label{Fried1}
\end{equation}
\begin{equation}
\frac{\dot{a}^2}{a^2}+2\frac{\ddot{a}}{a}+2\left(2a^{-6}\dot{a}^2-a^{-5}\ddot{a} \right)b=
-8\pi G\left[P-\Pi-\Pi b \right]
\label{Fried_dg2}
\end{equation}
\begin{equation}
\frac{\dot{a}^2}{a^2} +2\frac{\ddot{a}}{a} +\left(2a^{-5}\ddot{a}+a^{-6}\dot{a}^2 \right)b=
-8\pi G\left[ P+ba^{-4}P+2\Pi +2\Pi b \right]\label{Fried33}
\end{equation}
\begin{equation}
\left(\frac{\dot{a}^2}{a^2}+2\frac{\ddot{a}}{a}\right)b=8\pi G\left(Qa+\mu b \right)
\label{Fried_extra}
\end{equation}
where $b$ is considered constant and small. We have also made the approximations
$bQ, b\dot{Q}\approx 0$ since $q_\mu$ is of first order. This system of differential equations will be investigated with the aid of the linear dependent continuity equations.

The relation (\ref{Fried1}) is a modified Friedmann equation with an extra term $\Lambda(t)/3=4a^{-6}\dot{a}^2b/3$, analogous to $b$. The sign of $b$ will determine the effect of this extra geometrodynamical quantity, as we will briefly demonstrate in Sec.8. The term $\Lambda(t)/3$   acts as an ``effective'' time dependent cosmological constant, which tends fast to zero for an expanding universe. However, at earlier times this Lorentz Violating ``contribution'' can crucially  affect the dynamics.  In case $b$ varies at different geometrical phases, someone must include derivatives of $b$ to describe this general structure.

\section{The continuity equation}
The definition of the $u^\mu$-frame validates the conservation law $\nabla_{\nu}T^{\mu\nu}=0$, where the covariant derivative comes from the osculating metric $g_{\mu\nu}(x)$  (\ref{oscmetr}). The non-zero components of the energy-momentum tensor's divergence lead to a set of two differential equations; the time-like part provides the energy density formula
\begin{equation}
 \dot{\mu}+3\frac{ \dot{a} }{a}(\mu+P)\left(1-\frac{2}{3}a^{-4}b\right)=0\label{cont_full}
\end{equation}
and the space-like part the momentum density conservation
\begin{equation}
 \dot{Q}+3Q\frac{\dot{a} }{a}=-2 (P+\mu)a^{-2}\dot{a}b-(\dot{P}+\dot{\mu})a^{-1}b \label{contq}
\end{equation}
where all the quadratic terms of $b,Q$ have been dropped out.  The ordinary differential equation (\ref{cont_full}) gives back the  solution
\begin{equation}
\mu (t)=\mu_0 a(t)^{-3(1+w)}\exp\left( -\frac{1+w}{2}a(t)^{-4}b \right)\label{mu_solut_cont}
\end{equation}
where we have applied the equation of state $w=P/\mu$ and $\mu_0$ is an integration constant. After plugging (\ref{mu_solut_cont}) into (\ref{contq}), a direct calculation determines $Q$
\begin{equation}
Q(t)=-\mu_0(w+1)\exp\left[-\frac{1}{2}b(1+w)a(t)^{-4}  \right]a(t)^{-3w-4}b+Q_0 a(t)^{-3}.\label{qsolut}
\end{equation}
 In the special case $w=-1$, the energy density evolves as in the FRW model while the energy flux decays as  the  standard dust limit.

\section{Solutions for the scale factor and the anisotropic pressure }
We can recast the system of the ordinary differential equations (\ref{Fried1})-(\ref{Fried_extra}) to a more convenient form, using the linear dependent continuity equations (\ref{cont_full}),(\ref{contq}).
The solution for the scale factor $a(t)$ can be directly provided by the modified Friedmann equation (\ref{Fried1}). After substituting
the energy density (\ref{mu_solut_cont}) to (\ref{Fried1}) we derive
\begin{equation}
\frac{\dot{a}^2}{a^2}-\frac{4}{3}a^{-6}\dot{a}^2b=\frac{8\pi G}{3}\mu_0a^{-3(w+1) }\left( 1-\frac{w+1}{2} a^{-4}b    \right)+O(b^2)
\label{frdsol}
\end{equation}
where all the quadratic terms of $b$ have been omitted since we are interested in a first order approximation of the
unknown parameter. We present the analytical solutions of (\ref{frdsol}).
\newline\newline
{\it Solutions for $a(t)$, $w\neq -1$}\newline
The integration of (\ref{frdsol}) for $w\neq -1$ implies the solution
\begin{equation}
t\propto \frac{4}{w+1}a^{3(w+1)/2}+a^{(3w-5)/2}b.\label{asol1}
\end{equation}
The second term of (\ref{asol1}) reflects the contribution of the underline Finslerian  theory to the expansion dynamics. Given an expanding phase, the first term of (\ref{asol1}) increases faster compared to the second one, dominating the solution  . When $b$ is set to zero we recover the flat FRW solution.
\: \newline\newline
{\it Solutions for $a(t)$, $w= -1$}\newline
In the $w=-1$ case the Friedmann equation is simplified to the form
\begin{equation}
3\frac{\dot{a}^2}{a^2}-4a^{-6}\dot{a}^2b=8\pi G \mu_0 \label{frwm1}
\end{equation}
 The differential equation (\ref{frwm1}) is integrated to the logarithmic solution
\begin{equation}
t\propto 6\ln (a)+a^{-4}b.\label{asol2}
\end{equation}
indicating a rapid expansion.
 \newline\newline
{\it The anisotropic pressure $\Pi(t)$ }\newline
After calculating the scale factor $a(t)$ we can subtract
(\ref{Fried_dg2}),(\ref{Fried33}) and  derive the following expression for the anisotropic pressure $\Pi(t)$
\begin{equation}
 \Pi(t)=-\frac{1}{3}\mu_0 wa(t)^{-3(w+1)} b+\left( 8\pi G\right)^{-1}\left(a^{-6}\dot{a}^2-\frac{4}{3} a^{-5}\ddot{a} \right)b +O(b^2). \label{pitsol}
\end{equation}
  We illustrate the evolution of anisotropic pressure (\ref{pitsol}) for indicative values of $w$ at first order approach
  \begin{equation}
  \Pi(t)\propto \left\{
  \begin{array}{rll}
  \frac{b}{t^{14/3}}, & w=0, &\mbox{{\it matter}} \\
  \: & \: & \: \\
  -\frac{\mu_0}{t^2}b+\frac{21}{4t^4}b, & w=1/3, & \mbox{{\it radiation}} \\
  \: & \: & \: \\
  -e^{-\frac{2}{3}t}b, & w=-1, & \mbox{{\it dark energy}}
  \end{array}
     \right.
     \label{Pibeh}
  \end{equation}
  The anisotropic pressure fades out as $t$ grows up. Nevertheless, $\Pi(t)$ tends to infinity for ordinary matter at early times, while at the exotic case ($w=-1$) behaves exponentially.

  \section{Modified Potential for a zero energy particle}
The properties of the model can be further investigated using the zero energy particle approach.  The Friedman equation of motion (\ref{Fried1}) can be written in a form that represents the conservation of a particle's kinetic and potential energy. Substituting the energy density solution (\ref{mu_solut_cont}) in (\ref{Fried1}) we retrieve
\begin{equation}
\begin{array}{rl}
V(a)\propto & -\frac{a^{-3w-1}}{1-\frac{4}{3}a^{-4}b}\exp\left(-\frac{1+w}{2}a^{-4}b \right) \\
\: & \: \\
    = & -{a}^{-1-3w}+\frac{3w-5}{6}{a}^{-5-3w} b+O(b^2).
\end{array}\label{Vb}
\end{equation}
The first term of (\ref{Vb}) refers to the classical FRW-potential, while the second one is due to macroscopic consequences of  the  assumed local anisotropic structure. We restrict our investigation for $w<5/3$ which ensures that the behavior of the potential depends solely on the sign of $b$. consider,  an ingoing point particle falling into the potential (\ref{Vb}), that represents a collapsing universe. As we approach $a\rightarrow 0$, the kinetic energy of the particle diverges significantly from the FRW limit, since the $a^{-6}$ term dominates the dynamics; away from the initial singularity the LV effect fades out.
\begin{figure}
\begin{center}
\includegraphics[width=0.7 \textwidth, angle=0]{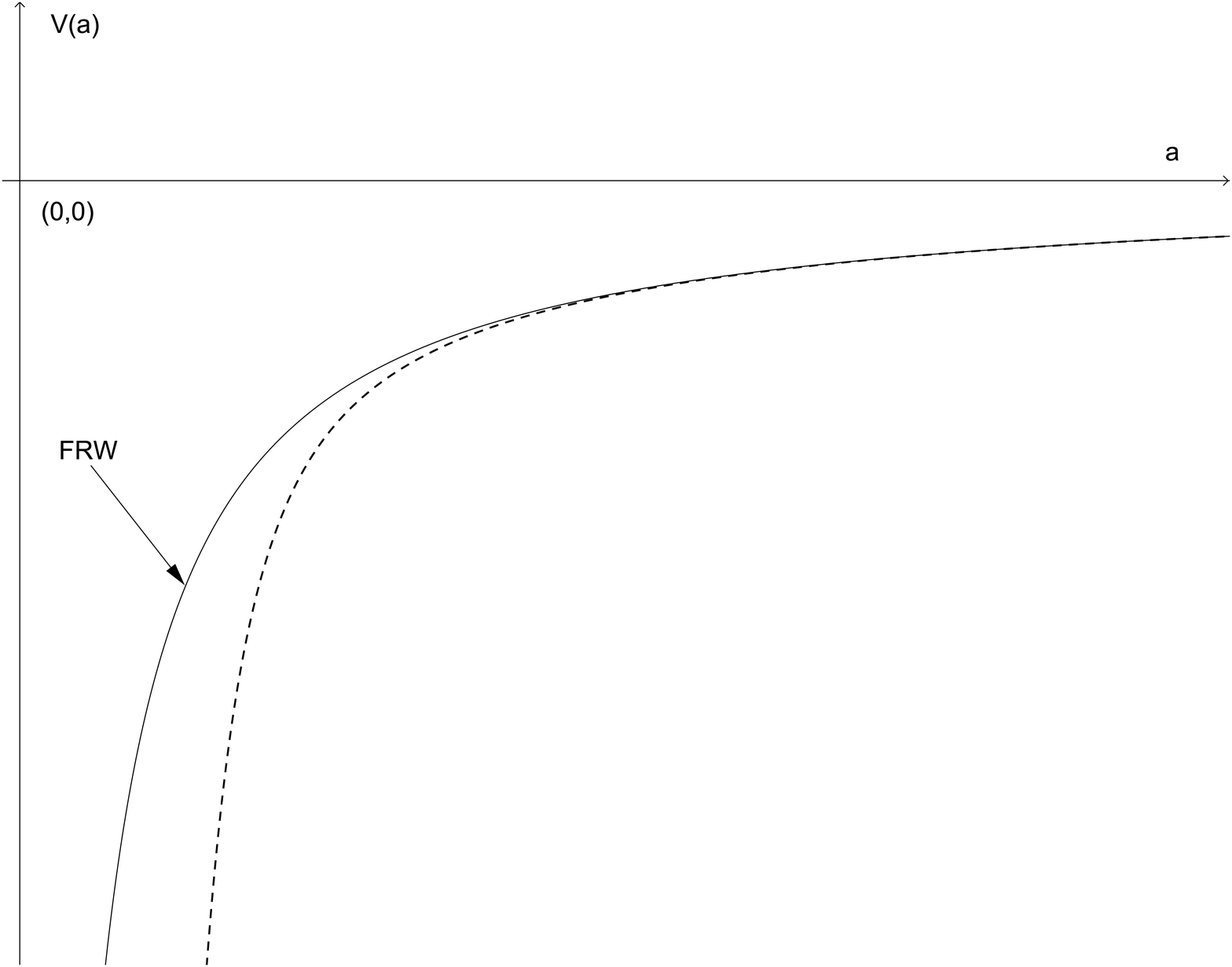}
\end{center}
\caption{\small Free scale potential for the analogous zero energy 1-dimensional dynamics with $b>0$, $w>-1/3$; the continuous line is the FRW potential $(b=0)$. Note that the model's potential approaches $-\infty$ faster than the classical one.}
\end{figure}

Positive values of $b$ guarantee that the point particle hits the initial singularity faster than the FRW case, since the potential decays at more rapid rate (see fig.1). On the other hand, negative values of $b$  assure an extremal point
\begin{equation}
a_*=6^{-1/4}\sqrt{\frac{(5-3w)(5+3w)}{1+3w}}(-b)^{1/4}\label{ast}
\end{equation}
where the accelerating contraction turns to  a decelerating phase for $-1/3<w<5/3$ (see fig.2). As the particle falls further down, it is finally bounced by the potential at the point $V(a_{bc})=0$,
\begin{equation}
a_{bc}=\left(\frac{5-3w}{6}\right)^{1/4}(-b)^{1/4}.
\end{equation}
When the particle reaches this turning point, the decelerating contraction is reverted to an accelerating expansion until it crosses the extremal $a_*$ again. The following decelerating phase recovers gradually the FRW model with $\Lambda =0$.
\begin{figure}
\begin{center}
\includegraphics[width=0.7 \textwidth, angle=0]{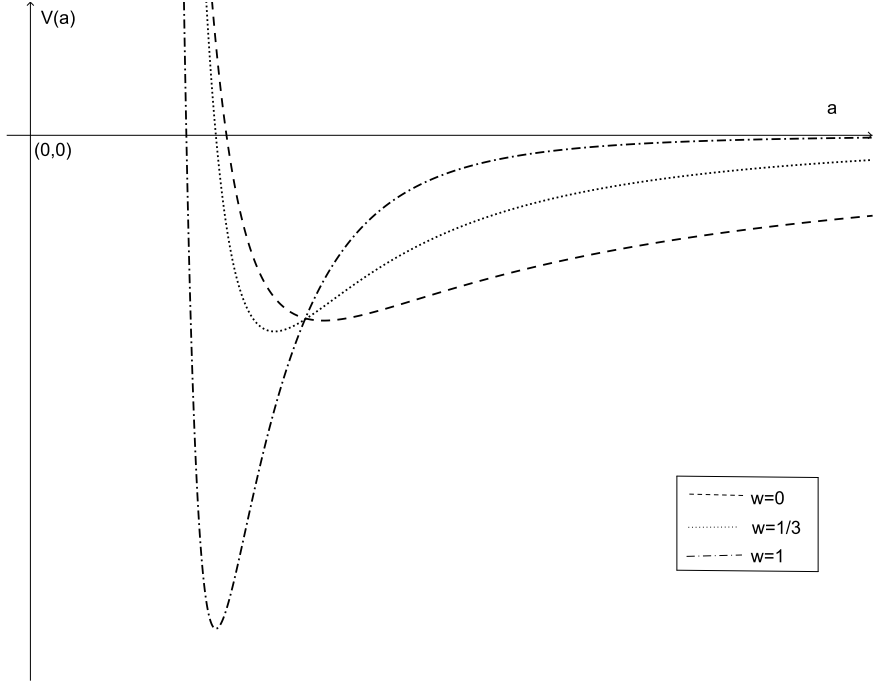}
\end{center}
\caption{\small Free scale potential for $b<0$. The zero energy particle approaching from $a\rightarrow\infty$ accelerates until it crosses the minimum of the potential $V(a_*)$; then decelerates until it hits the potential at the bouncing point. The model recovers FRW behavior for large values of $a$. }
\end{figure}
Using the solution (\ref{asol1}) we can roughly estimate the duration of the  accelerating phase between the bouncing and the extremal point
\begin{equation}
\Delta t\sim\left(\frac{8\pi G}{3}\mu_0\right)^{-1/2} (-b)^{3(w+1)/8}
\end{equation}
where $\mu_0$ is the energy density for $t=t_0$. If $\mu_0$ corresponds to the present matter density distribution, close to the critical density $\mu_{cr}\approx 10^{-29} gr\cdot cm^{-3}$, then approximately  $\Delta t\approx 10^{14}sec $ for the {\it ether drift experiment} and $\Delta t \approx 10^8 sec$  for the {\it anisotropy of inertia}. Thus, according to estimations of the age of old stars $\sim 3\times 10^{17}$ \cite{age01} our model, starting from $a_{bc}$, has entered a decelerating phase.  This conclusion is not in alliance with the observed cosmic acceleration
from the Type 1a supernova data. Exotic matter with $w<-1/3$ must be assumed for reproducing the desired expanding behavior. An alternative way to introduce late time acceleration in the present epoch is the adoption of the present energy density distribution to lower values. In particular, the values, $\mu_0\sim 10^{-32} gr\cdot cm^{-3}$ for {\it ether drift experiment} and $\mu_0\sim 10^{-42}gr\cdot cm^{-3}$ for the {\it anisotropy of inertia} secures that the model accelerates until today. In case of  clusters that are large enough to be representatives of the overall mass density,  only the limit of the {\it ether drift experiment} is close, yet less than the up to date density measurements \cite{Bahcall:1998ek}.
\section{Investigation of kinematical quantities}
The 4-velocity $u^\mu$ of the observer's rest frame introduces a ``slit'' between space and time. The vector $u^\mu$ determines the projective tensor $h_{\mu\nu}$, which acts as the  3-dimensional metric of the observer's instantaneous rest space for a hypersurface orthogonal congruence.

  Let a congruence  $x^{\mu}(\tau)$ consisted of non-geodesics, where the tangent vector field is the 4-velocity  $u^\mu=(1,0,0,0)$ and the line element (\ref{line_elem}) implies the shear vorticity and expansion
tensors.
The tensor field
\begin{equation}
\nabla_{\beta}u_\alpha= \frac{1}{3}\theta h_{\alpha\beta}+\sigma_{\alpha\beta}+\omega_{\alpha\beta} +A_{\alpha}u_{\beta} \label{decom}.
\end{equation}
measures the failure of deviation vector $\xi^\alpha$ to be parallel transported along the congruence.  The
expansion scalar $\theta$ is defined as $\theta= \nabla^\alpha u_\alpha$
and measures the expansion of a volume element $dV$ along $u^\mu$,  the shear tensor
$\sigma_{\alpha\beta}=\nabla_{\langle\beta}u_{\alpha\rangle}=h_{( \alpha}^{\:\:\rho}h_{\beta )}^{\:\:\sigma}\nabla_{\rho}u_{\sigma}-\frac{1}{3}\theta h_{\alpha\beta}$
represents deformations in the shape of $dV$, while the vorticity tensor
$\omega_{\alpha\beta}=\nabla_{[ \beta}u_{\alpha ]}$
expresses orientation changes of $dV$. After expanding for small values of $b$ we can derive the following approximation series
\begin{equation}
\theta=3\frac{\dot{a}}{a}-2\dot{a}a^{-5}b+O(b^2) \label{th}
\end{equation}
and
\begin{equation}
\sigma_{\alpha\beta}=\mbox{diag}\left[0,\:-\frac{2}{3}a^{-3}\dot{a},\:-\frac{2}{3}a^{-3}\dot{a},\:\frac{4}{3}a^{-3}\dot{a}\right]b+O(b^2)
\end{equation}
while
\begin{equation}
\omega_{\mu\nu}=0.
\end{equation}
Thus the hypersurface orthogonal condition holds for the congruence. Additionally, The non-vanishing sheer tensor reflects the kinematical anisotropies of the fluid.  An interesting point is the presence of the acceleration vector
$A^\mu=u^\nu(\nabla_\nu u^\mu)$
with the non-zero component
\begin{equation}
 A^\mu=(0,0,0,a^{-4}\dot{a})b+O(b^2)\:.
\end{equation}
The presence of  acceleration represents non-gravitational phenomena.  Therefore, the fundamental observer is not moving along geodesics since Lorentz violations contribute substantially to the gravitational theory. This is a direct result from the distortion of the line-element (\ref{DS}), coming from the $b$ parameter.

\section{Raychaudhuri equation and focusing theorem}
The fluid's volume evolution is retrieved by the time-like part of the Ricci identities which lead to the well known Raychaudhuri equation. In case of a hypersurface orthogonal geodesic congruence and ordinary matter the Raychaudhuri        equation implies the formation of a singularity due to the attractive nature of gravity. However,  a non-geodesic congruence may avoid the caustic since the external forces may resist the collapse. Given a non-geodesic congruence, Raychaudhuri equation reads
\begin{equation}
\frac{d\theta}{d\tau}= -\frac{1}{3}\theta^2-2(\sigma^2-\omega^2)-R_{\mu\nu}u^\mu u^{\nu}+D^{\mu}A_{\mu}-A_\mu A^\mu \label{raych}
\end{equation}
where $D_{\mu}A_{\nu}=h_{\mu}^{\:\:\alpha}h_{\nu}^{\:\:\beta}\nabla_{\alpha}A_{\beta}$ denotes the spatial gradient of a vector.

The line element (\ref{line_elem}) validates the hypersurface orthogonality of the $u^\mu$-congruence, $\omega_{\mu\nu}=0$. Thus, the only quantities that may offer a positive contribution in the rhs of (\ref{raych}) are related to 4-acceleration $A^\mu$.
In particular, the square magnitude of the 4-acceleration always resists the collapse (assists the expansion) since it is a space-like vector, while $D^{\mu}A_{\mu}$ depends on the state of the expansion. The quantity $D^{\mu}A_{\mu}$ keeps always a positive sign for a universe at a decelerating phase. This positive sign is still ensured for an accelerating phase if $\ddot{a}<4\dot{a}a^2$. The effect of Lorentz violations  wins over the attractive nature of gravity if the following condition holds
\begin{equation}
D^{\mu}A_{\mu}-A_\mu A^\mu >R_{\mu\nu}u^\mu u^{\nu}+2\sigma^2
\end{equation}
 A first order approximation of the additional terms in (\ref{raych}) imply
$D^{\mu}A_{\mu}-A_\mu A^\mu =O(b^2)$. However, at an earlier stage of the expansion,
where the Lorentz violations are stronger, the
accelerating terms may dictate over the negative ones preventing the collapse of the fluid to a geometrical singularity. On the other hand, if the values of $b^2$ are
comparable to the cosmological constant, the $D^{\mu}A_{\mu}-A_\mu A^\mu$-term may give rise to the dark energy scenario of an almost empty self-accelerating universe.

\section{Lorentz violations as a source of inhomogeneities}
We consider a space-time where the universe is regarded as a single imperfect fluid, a direct result of the process of osculation. Spatial inhomogeneities
in the cosmic fluid are detected by the following dimensionless, gauge-invariant quantities \cite{EllBr, StWk}
\begin{eqnarray}
\Delta_{\alpha}=\frac{a(t)}{\mu}D_\alpha \mu \label{deltas}\\
\mathcal{Z}_\alpha =a(t)D_\alpha \theta  \label{zetas}
\end{eqnarray}
 which both vanish in spatially homogeneous space-times. The tensors (\ref{deltas}),(\ref{zetas}) do not vanish even in case they are zero for a specific value of the time coordinate.  They
are considered as some of the key sources of the density perturbations \cite{tsagas_rep}. Despite the solely dependence on time of $\mu$ and $\theta$ their spatial gradient does not vanish since $h_a^{\:\: 0}\neq 0$. The evolution of (\ref{deltas}),(\ref{zetas}) at first order are described by the differential equations \cite{tsagas_rep} \footnote{$\dot{\Delta}_{<\kappa>}=h_\kappa^{\:\:\rho}\dot{\Delta}_\rho =h_\kappa^{\:\:\rho} u^\nu\nabla_\nu\Delta_{\rho} $}
\begin{equation}
\dot{\Delta}_{<\kappa>}-\frac{P}{\mu}\theta\Delta_{\kappa}+\left(1+\frac{P}{\mu}    \right)\mathcal{Z}_{\kappa}+\sigma ^{\nu}_{\:\:\kappa}
\Delta_{\nu}= \frac{a\theta}{\mu}\left( \dot{q}_{<\kappa>}+\frac{4}{3} \theta q_{\kappa} +\sigma_{\kappa\lambda}q^\lambda \right)
\label{ddelt}
\end{equation}
and
\begin{equation}
\dot{ \mathcal{Z} }_{<\kappa>}+\frac{2}{3}\theta\mathcal{Z}_{\kappa}+
+\frac{1}{2}\mu\Delta_\kappa+\frac{3}{2}aD_\kappa P = -a\left[\frac{1}{3}\theta^2+\frac{1}{2}(\mu-3P)  \right]A_\kappa\label{dzeta}
\end{equation}
where flux and acceleration act as sources of  perturbations. Hence, this mechanism indicates that Lorentz violations may generate inhomogeneities. A direct calculation gives back the non-vanishing components
\begin{equation}
 \Delta_3=3b(1+w)\frac{ \dot{a} }{a}+O(b^2),\:\:\: \mathcal{Z}_3=-3b\left(\frac{\ddot{a}}{a}-\frac{\dot{a}^2 }{a^2} \right)+O(b^2).\label{DZ}
\end{equation}
 where $a(t)$ is the FRW solution since the leading terms of (\ref{DZ}) are proportional to $b$. Note that, in  case of $w=-1$ the density inhomogeneities are zero.

\section{Discussion}
The osculation of a Finslerian manifold generates a Riemannian cosmological {\it toy model} for a specific 4-velocity $u^\mu$.   The rest frame of the fundamental observer lies on a tilted non-geodesic congruence. The peculiar velocity $u^\mu$  defines the energy-momentum tensor of an imperfect fluid, given an almost  FRW-metric of zero spatial curvature. This construction holds for  $u^{\mu}=(1,0,0,0)$ to retrieve the standard FRW limit of the comoving observer, in case of vanishing Lorentz violations. The resulting line element of  space-time describes an anisotropic expanding (contracting) medium. Therefore, an imperfect fluid with flux and anisotropic pressure is ``injected '' into space to support this anisotropy.

 The field equations lead to a  modified  Friedmann equation of motion with an effective varying cosmological constant proportional to $b$. The spatial curvature is considered $k=0$ in accordance to the astrophysical observations. The Einstein field equations combined with the conservation laws for the flux and energy density, provide analytical solutions for the scale factor, anisotropic pressure, energy density and flux. However, the expressions for the scale factor are given in the form $t\equiv t(a)$. The model's differential equations add a first order exponential factor to the FRW's standard energy density while the flux and anisotropic pressure tend to zero for an increasing $a(t)$. We remark that the flux vector does not directly vanish even if the parameter $b$ is set to zero.

 Furthermore, we investigate the model's behavior using the zero energy particle with a modified potential, in relation to the FRW case. The parameter $b$ plays an influential role, since it determines substantial properties of the dynamics. In particular, for $b>0$ a contracting universe will hit the initial singularity faster than the FRW. On the other hand, if $b<0$ a richer scenario occurs; an accelerating contraction leads to the potential's local minimum, turning to a decelerating contraction until it bounces back at some turning point. After the particle's reflection to the potential, an accelerating phase takes place until we cross the extremal point. Finally, a decelerating expansion governs the evolution due to the attractive nature of gravity, recovering the FRW limit with $\Lambda=0$. In order to reproduce late time acceleration two possibilities rise up: assume exotic matter with $w<-1/3$, or a `lighter' universe with $\mu_0<\mu_{crit}$ today. Nevertheless, the time dependent effect to the kinematics, coming from the LV parameter,  gives birth to a direct discrimination to the  $\Lambda CDM$ structure since the extra term can be addressed as an ``effective'' varying cosmological constant.

 One key  geometrical quantity to ``translate'' observational data is the luminosity distance $d_L(a)$. The  form of this function depends on the physical hypotheses taken into account. In that sense, the interpretation of the  observations is model dependent. Violating the symmetries of space-time in large distances, $d_L(a)$ might be crucially  affected. A first question is how  sensitive  this function can be in terms of any extra parameters introduced by the model.  If our model is not sensitive to the parameter $b$, one could accurately determine whether or not  the effective varying cosmological constant plays a role to the expansion dynamics. Even in such a case, observations suggest that a constant $\Lambda$-term is completely consistent to the current data. On the other hand, due to many theoretical motivations, there has been an extensive amount of work investigating whether improved observational scenarios could measure a  time-dependent $\Lambda$  behavior. These surveys are mainly directed by Baryon  Acoustic Oscillations, SNIa, Lensing  and CMBR data \cite{Ltrev}.

 In GR ordinary matter always falls along geodesics at the presence of gravity alone. The study of gravitational collapse leads to the  formulation of singularity theorems. However, other phenomena in nature impose non-geodesic motion (for a recent review see \cite{tsagas_rep}). In this arena the question of a caustic singularity  must be revisited. A characteristic example is the collapse of an ideal MHD fluid where magnetic tension may prevent the formation of a caustic \cite{Tsagas_grav_collapse}. The present phenomenological model points out that the osculation of a Finsler space to a Riemannian one leads to non-geodesic motion.  The observer's 4-acceleration vector mimics Lorentz violations for our effective theory. Therefore, Raychaudhuri's equation implies that  the formation of a singularity depends on the magnitude of the parameter $b$ which gives birth to Lorentz violations.

 The presence of flux at the energy-momentum tensor is a direct consequence of the reduced non-comoving motion, since $g_{0i}\neq 0$. The existence of flux combined to the peculiar motion acts as a source of inhomogeneities. As a result, relics of Lorentz violations are encoded to density perturbations. In the context of Finsler geometry, this mechanism  demonstrates a possible correlation of Lorentz violations to CMB physics, pointing out some possible future developments in the field.

 A further research would be the construction of a spatially curved modified FRW model and the back-reaction of  Lorentz violations to spatial curvature. Also, the concept of geometrical phase transitions via Finslerian geometrical structures is of some interest. In the framework of GVSR we can express these phase transitions by setting $b$ dependent on the space-time coordinates. The study of the $b$-evolution may shed light to the question ``{\it why $b$ is so small?}''. Finally, a vital task still remains: how a consistent modified gravitational theory can be achieved using the whole machinery of Finsler geometry \cite{Vacaru:2007ng}.

\section{Acknowledgments}
  The authors would like thank C.G.Tsagas for the useful discussions on non-geodesic motion and the University of Athens (Special Accounts for Research Grants) for the support to this work.

\end{document}